\documentclass{webofc}
\usepackage[varg]{txfonts}   % Web of Conferences font
\usepackage{hyperref}
\usepackage{url}

\usepackage{graphicx}
\usepackage{subfig}
\usepackage{gensymb}

\hypersetup{colorlinks=true,citecolor=blue,urlcolor=blue,linkcolor=blue}
\begin{document}
\title{Next Generation LLRF Control Platform for Compact C-band Linear Accelerator}

\author{\firstname{Chao} \lastname{Liu}\inst{1}\fnsep\thanks{\email{chaoliu@slac.stanford.edu}}\and
        \firstname{Ryan} \lastname{Herbst}\inst{1} \and
        \firstname{Larry} \lastname{Ruckman}\inst{1}\and
        \firstname{Emilio} \lastname{Nanni}\inst{1}\
        % etc.
}

\institute{SLAC National Accelerator Laboratory, Menlo Park, California, USA 
          }

\abstract{%
 The Low-Level RF (LLRF) control circuits of linear accelerators (LINACs) are conventionally realized with heterodyne based architectures, which have analog RF mixers for up and down conversion with discrete data converters. We have developed a new LLRF platform for C-band linear accelerator based on the Frequency System-on-Chip (RFSoC) device from AMD Xilinx. The integrated data converters in the RFSoC can directly sample the RF signals in C-band and perform the up and down mixing digitally. The programmable logic and processors required for signal processing for the LLRF control system are also included in a single RFSoC chip. With all the essential components integrated in a device, the RFSoC-based LLRF control platform can be implemented more cost-effectively and compactly, which can be applied to a broad range of accelerator applications. In this paper, the structure and configuration of the newly developed LLRF platform will be described. The LLRF prototype has been tested with high power test setup with a  Cool Cooper Collider (C\(^3\)) accelerating structure. The LLRF and the solid state amplifier (SSA) loopback setup demonstrated phase jitter in 1 s as low as 115 fs, which is lower than the requirement of C\(^3\). The rf signals from the klystron forward and accelerating structure captured with peak power up to ~16.45 MW will be presented and discussed.
}
\maketitle
\section{Introduction}
\label{intro}

The low-level radio frequency (LLRF) system typically controls the electromagnetic field in accelerator cavities with extremely high precision to achieve the optimum beam quality. For accelerators operating in the GHz range, the convectional LLRF control circuits are implemented with heterodyne based architectures, which have analogue up and down mixers to convert base-band signal to and from RF frequency ranges. The RF pulses from the cavities are down mixed to base-band and  digitized by discrete ADCs. The digital data is then streamed to an FPGA and processed by the feedback algorithm implemented within it. Based on experiment and user requirements, updated base-band pulses are calculated in FPGA and generated by discrete DACs and up converted by RF mixers. The AMD Xilinx RFSoC device integrates the majority of the essential components for a LLRF mentioned above in a single system-on-chip (SoC). The direct RF sampling technique of RFSoC eliminates the RF mixing components and offers a flexible, compact and low power solution for LLRF controls system for rf cavities up to 6 GHz.

We have investigated RF performance of the RFSoC devices in different frequency ranges and developed readout and control platforms for a number of physics experiments \cite{liu2021characterizing, liu2022development,  liu2023evaluating, henderson2022advanced, liu2023higher,liu2024development}. For the accelerator applications, the new proposed lepton collider Higgs factory, Cool Cooper Collider (C\(^3\)), was the initial targeted application for the prototype RFSoC based LLRF \cite{Bai:2021rdg}. C\(^3\) has been proposed as compact and affordable machine and an RFSoC-based LLRF with high integration level can help to achieve the design goals \cite{Vernieri:2022fae}.  Based on the requirements of C\(^3\), we have performed a range of targeted configuration optimizations and performance characterizations and summarized the test results in \cite{liu2024direct}. The RFSoC evaluation board with our custom firmware designs demonstrated pulse to pulse phase jitter as low as 87.54 femtoseconds (fs), which is considerably better than the 150 fs requirement of C\(^3\) \cite{nanni2023status}. 

In \cite{liu2024direct}, the performance evaluation results were focused on the data converters and lower power levels. The test setup for evaluation were direct loopback from the DAC to ADC or via a custom designed solid state amplifier (SSA) with peak power up to 300 W. With the optimized RFSoC configurations, we have designed and implemented a new RFSoC based LLRF platform, named next generation LLRF (NG-LLRF). In this paper, the architecture of the NG-LLRF will be introduced and preliminary high power tests performed with C\(^3\) accelerating structure cavities using the test facility at Radiabeam will be summarized and discussed. 

\section{NG-LLRF System and Hardware}

\begin{figure}[h]
\centering
\includegraphics[width=1\columnwidth]{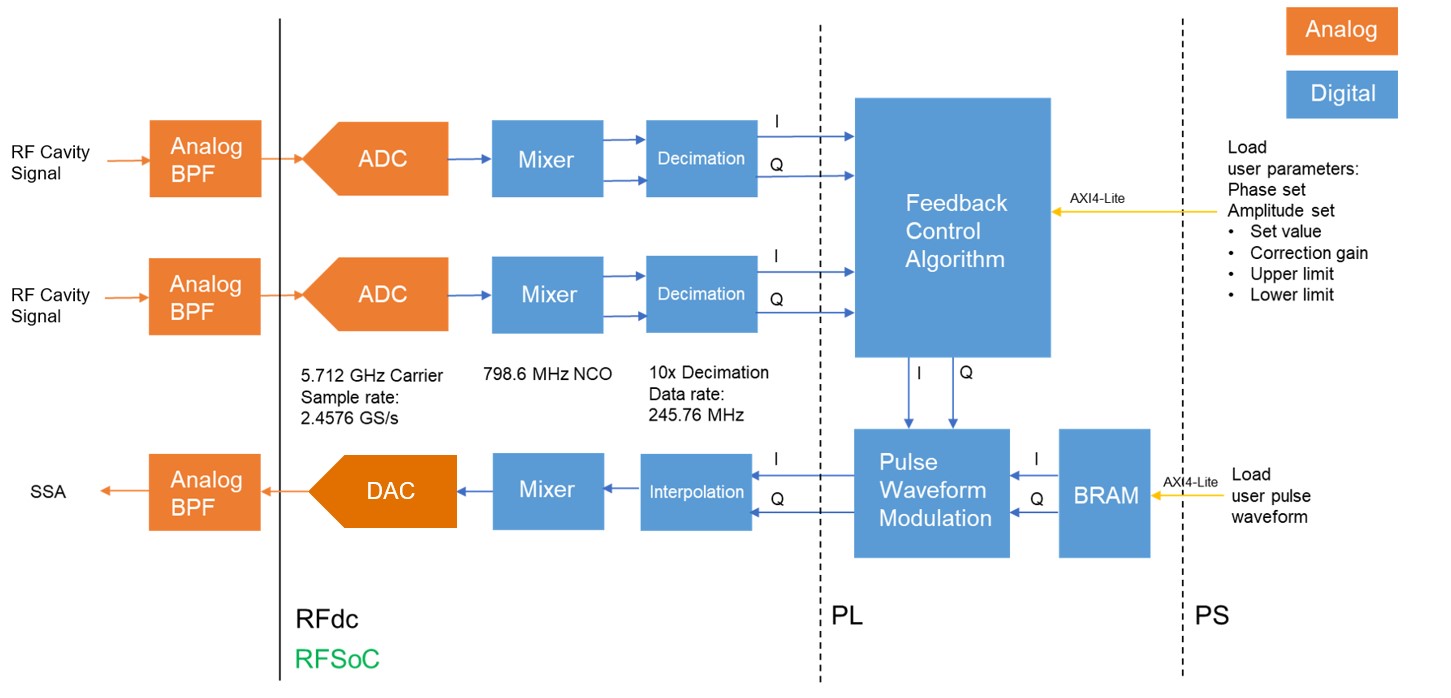}
\caption{ The block diagram of the new generation of LLRF control circuit for C-band LINAC. All the components or firmware blocks on the right side of the solid line are integrated or implemented in RFSoC. }
\label{fig-1}       % Give a unique label
\end{figure}

Figure \ref{fig-1} shows the block diagram of the NG-LLRF with two rf input channels and a single rf output channel for a single set of accelerating structure cavities. However,the NG-LLRF system can have up to 16 rf input channels and 16 rf output channels, which can be customized per application. 

The feedback loop is implemented with the three key modules in RFSoC, which are the processor system (PS), the programmable logic (PL) and the rf data converters (RFdc). The rf signals from the cavity are injected to the integrated ADCs of the RFSoC via band-pass filters (BPF). As it has been discussed in \cite{liu2024direct}, the BPFs are critical to achieve the optimum noise performance for direct rf sampling especially at higher order Nyquist zones. 

The center frequency of NG-LLRF is around 5.712 GHz with a bandwidth of 245.76~MHz. The ADCs sample the filtered rf signals with sampling rate of 2.4576 giga samples per second (GSPS), so the C-band rf frequency is in the fifth order Nyquist zone of the ADCs. The digital samples are down mixed digitally at the image of the centre frequency at first Nyquist zone. The baseband output of the mixer in in-phase (I) and quadrature (Q) format is then decimated and filtered. The feedback control algorithm takes IQ components of the down-converted cavity signal and calculates a new set of IQ values based on user input parameters. The updated IQ is modulated with the user defined base-band pulse in PL. The modulated pulse is then interpolated and up converted. The updated rf pulse is generated by the integrated DAC in RFSoC at sampling rate of 5.89824 GSPS with 5.712 GHz in the second Nyquist zone. For NG-LLRF users, all the configurable parameters or pulse waveform loading are handled in software layer implemented in the processing system (PS) integrated in RFSoC and accessible from the server connected via Ethernet. The user configurations and data capturing are performed with Jupyter Notebook and the software interface is also compatible with Experimental Physics and Industrial Control System (EPICS) process variables (pv). 

\begin{figure}[h]
\centering
\includegraphics[width=1\columnwidth]{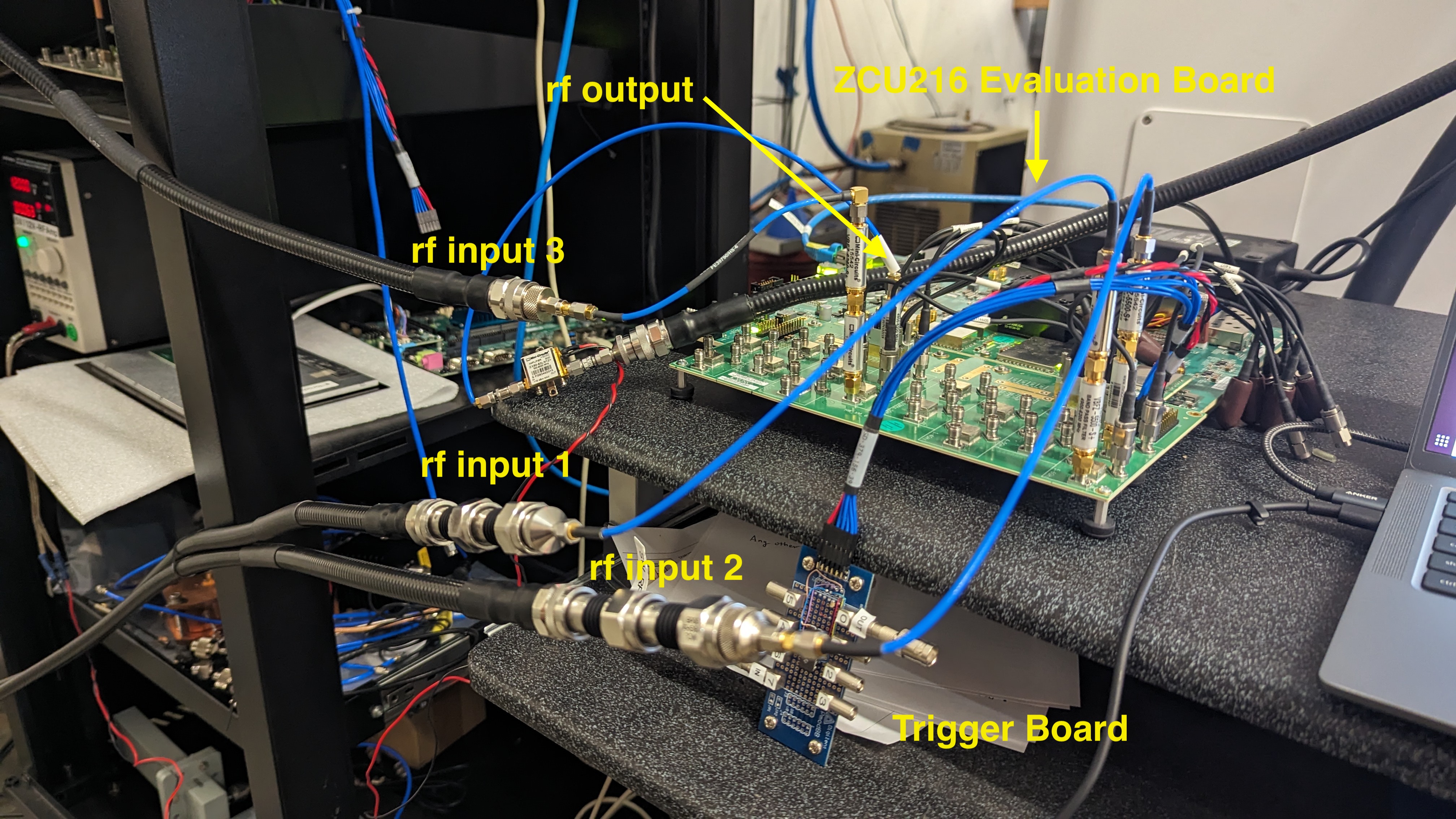}
\caption{ The hardware of the NG-LLRF with three RF input channels and one RF output channel. }
\label{fig-2}       % Give a unique label
\end{figure}

 Figure \ref{fig-2} shows the hardware setup of the prototype of NG-LLRF. The prototype is developed based on a ZCU216 evaluation board with an UltraScale+ XCZU49DR-2FFVF1760 RFSoC \cite{zcu216}. As Figure \ref{fig-2} shows, the platform has three rf inputs and one rf output. The rf pulse generated by the rf output is amplified by an SSA and then drives the klystron. The high power rf pulse from klystron is coupled to the C\(^3\) cavity structure in a test bunker via a waveguide. The resonance frequency of the C\(^3\) structure is approximately 5.71209 GHz, which is the rf frequency used for this exercise. The rf signals from the couplers at multiple stages in the test structures are looped back to the rf input channels of the NG-LLRF platform for measurement. For this NG-LLRF experiment, the maximum peak power delivered to the accelerating structure was as high as 16.45 MW. Attenuators are inserted to reduce the power level of signals down to 0 dBm maximum, which is below the 1 dBm maximum input power of ADCs in RFSoC.
 
 The pulse generation of rf output and data capturing at rf inputs are all triggered by an external TTL signal which is injected to the trigger board first and then propagate to the entire platform. The external trigger source used also triggers the operation of SSA and klystron. However, the NG-LLRF can also be configured as the master trigger source and delivering multiple trigger signals via the same trigger board. 

\section{High Power Test of NG-LLRF with C\(^3\) Accelerating Structure }

\subsection{SSA Output Characterization}

The SSA integrated in the klystron is typically driven by a commercial LLRF system. In the high power test, the DAC integrated in RFSoC of NG-LLRF will the drive of the entire test structure and the ADCs will measure the RF signals at multiple stages of the structure. Therefore, the RF performance of SSA with NG-LLRF in loopback mode is characterized first. In this test, the NG-LLRF is configured to generate rf pulses with a width of 1 \(\mu\)s at 60 Hz. The power level is adjusted by loading the baseband pulse with a different amplitude and the up mixing circuit shown in \ref{fig-1} will generate the rf pulse with the integrated DAC in RFSoC. The test has been performed with DAC amplitude from 2000 to 10000 with a step size of 2000, which corresponds to SSA output peak power level from 3.2 W to 60.8 W measured by a peak power meter. The SSA output is looped back to one of rf input channels of the NG-LLRF platform. The RF signal is digitized by the ADC and processed with down conversion chain. The base-band data for consecutive pulses is captured in IQ format and processed as needed for further analysis. 

\begin{figure}[!tbp]
  \centering
  \subfloat[The time domain plot of the SSA output.]{\includegraphics[width=0.5\textwidth]{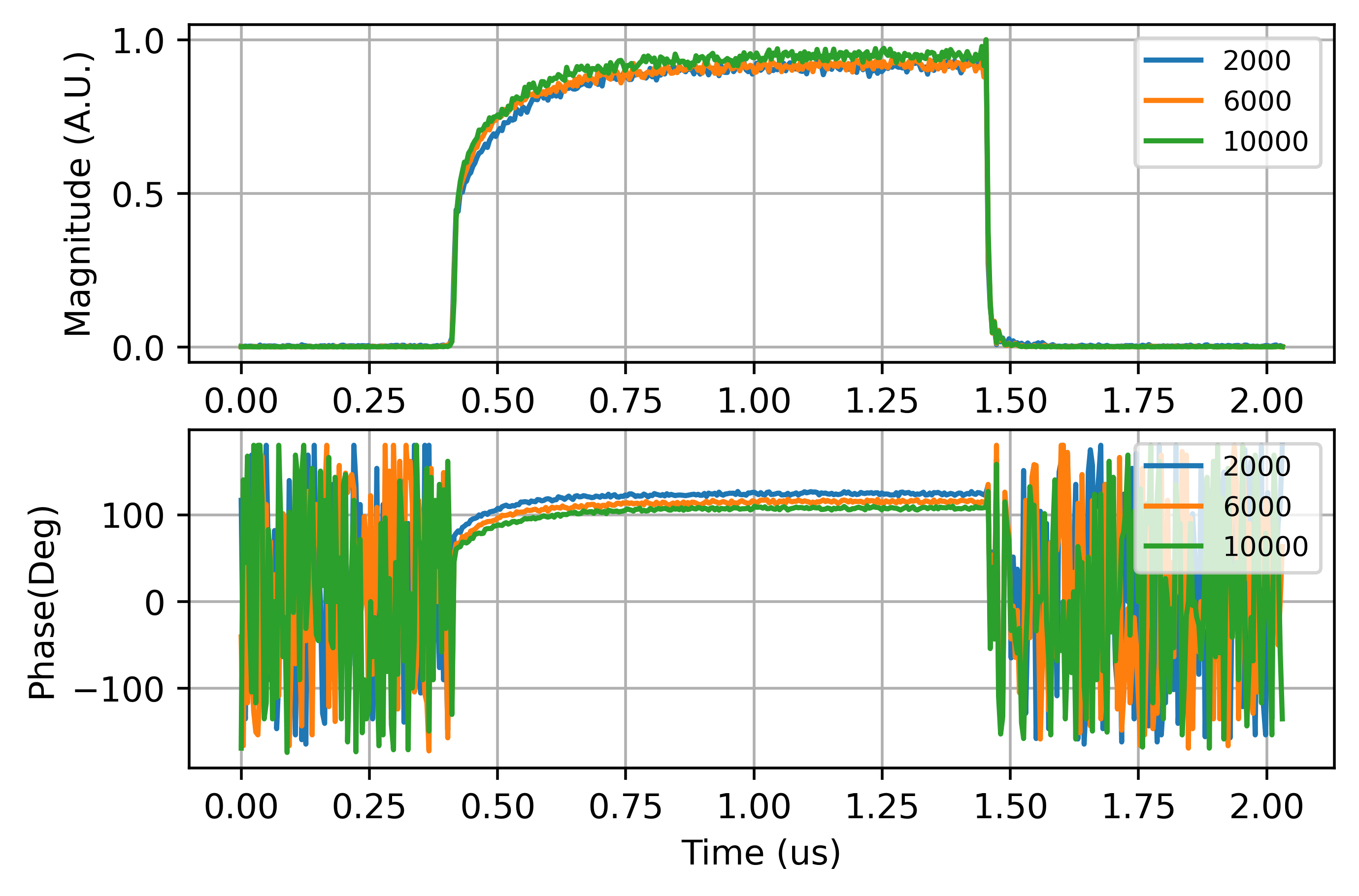}\label{fig:f3a}}
  \hfill
  \subfloat[The frequency domain plot of the SSA output on the flat top of the pulse.]{\includegraphics[width=0.5\textwidth]{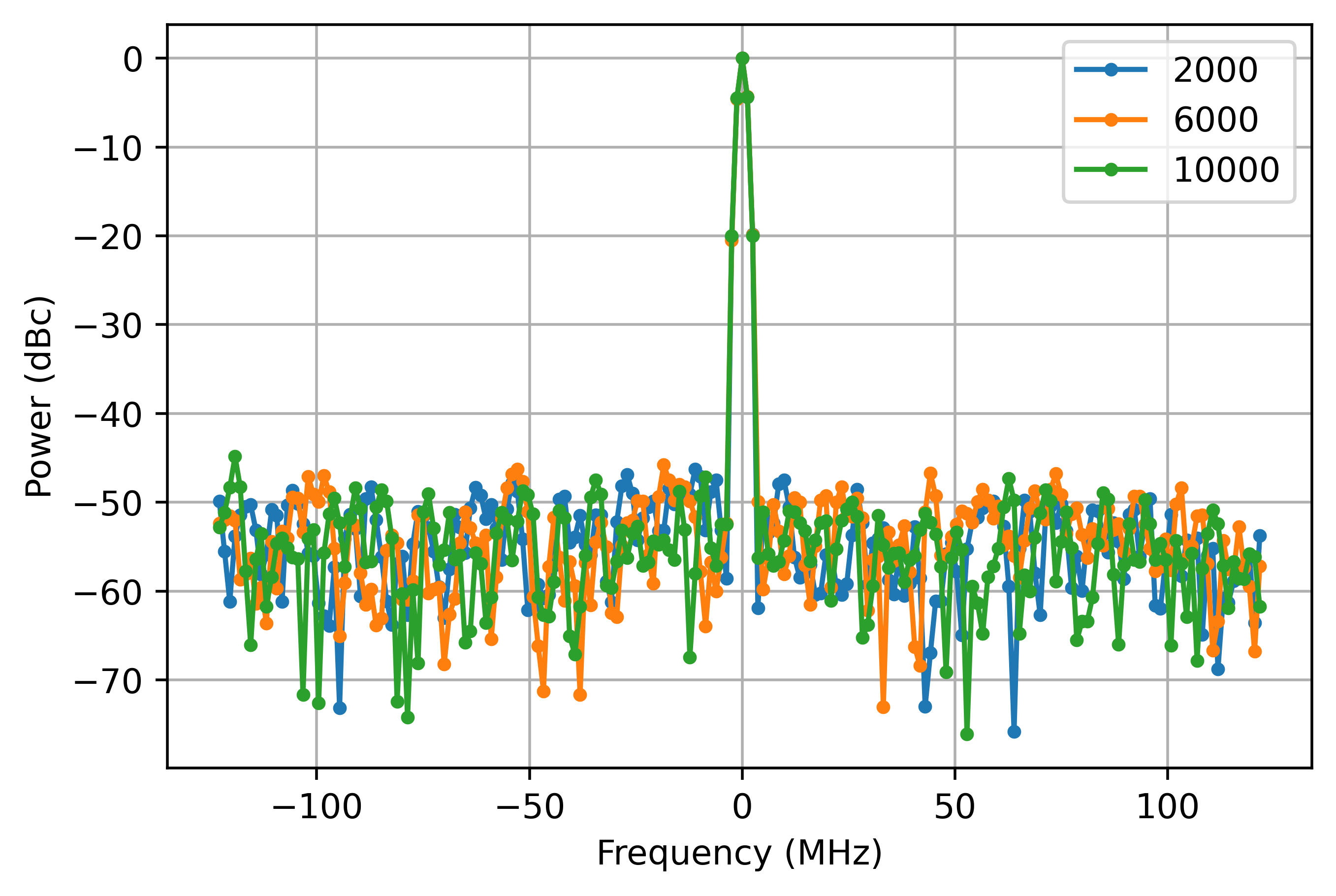}\label{fig:f3b}}
  \caption{The plots of SSA output signals via a attenuators at different DAC amplitude levels.}
\end{figure}

Figure \ref{fig:f3a} shows the time domain plot of the SSA output pulses at different DAC amplitude values. The magnitude values are scaled to the maximum of each measurement and rise time is approximately 0.3 \(\mu\)s, which similar for all three DAC amplitude levels. The phase value ramps are consistent for different DAC amplitude levels, but reduced by approximately 10 degrees of each 4000 increase in DAC amplitude values. The spectrum of the SSA output is calculated with 200 samples on flat top of pulse with same data set as above and shown in Figure \ref{fig:f3b}. The spectrum for DAC amplitude levels has similar profile. The spectrum is dominated by the tone at carrier frequency and there is no obvious tones with in the 245.76 MHz bandwidth, which are desirable as the drive of the klystron.

\begin{figure}[!tbp]
  \centering
  \subfloat[The average magnitude and phase values on the flat top of 60 consecutive SSA output rf pulses.]{\includegraphics[width=0.52\textwidth]{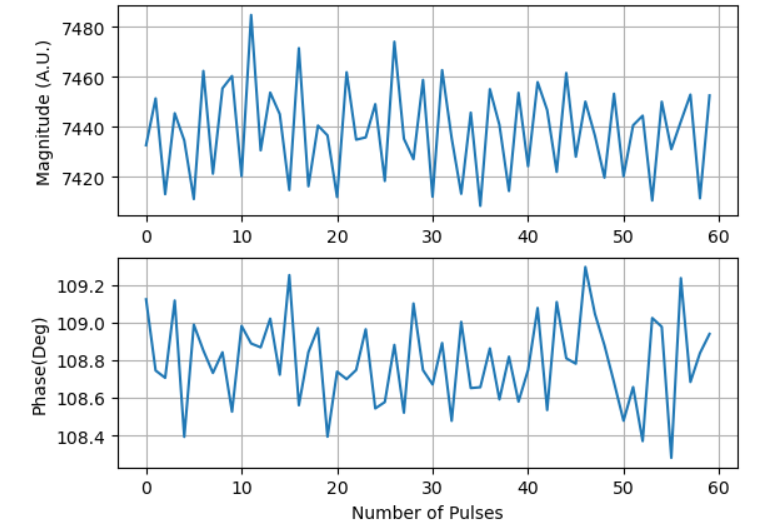}\label{fig:f4a}}
  \hfill
  \subfloat[The phase jitter measure at different power levels.]{\includegraphics[width=0.48
\textwidth]{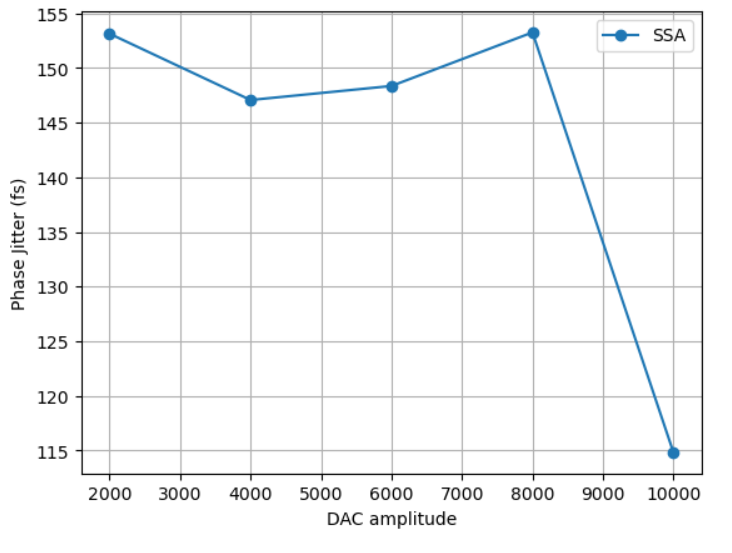}\label{fig:f4b}}
  \caption{The magnitude and phase fluctuation characterization on the flat top of SSA output signals.}
\end{figure}

For LINACs, the pulse to pulse magnitude and phase fluctuation measurements are critical performance indices. In this evaluation, the inter pulse fluctuation is evaluated with 60 consecutive pulses captured in 1 second. The average magnitude and phase values on the flat top is calculated for each of pulse and the standard deviation of the average values for the 60 pulses are used to evaluate the fluctuation levels. Figure \ref{fig:f4a} shows the plots of average magnitude and phase values for 60 pulse captured with the DAC amplitude at 10000. The standard deviation of magnitude values is approximately 0.23\% respect to the average magnitude of all the pulses. The standard deviation of the phase is around 0.24\degree. The phase jitter with unit of degree is converted to time units at the corresponding rf frequency. The plot in Figure \ref{fig:f4b} shows the phase jitter levels with different DAC amplitude levels. The phase jitter with DAC amplitude at 8000 or lower is fluctuation around 150 fs and gets as low as approximately 115 fs with DAC amplitude at 10000, which corresponds to the SSA with output peak power of 60.8 W. That indicates that SSA has better phase stability at higher power levels, which explains the 87.54 fs phase jitter measured in \cite{liu2024direct} at peak power over 200 W.

\subsection{High Power Test Analysis}

In the higher power test, the rf output channel of the NG-LLRF is driving the entire test structure and three rf signals are looped back to the rf input channels, including klystron forward (FWD), cavity forward (FWD) and cavity reflection (REF). The peak rf power delivered to the structure ranges from 4.21 MW to 16.45 MW.  The LLRF system aims to measure, regulate and stabilize the rf field in the accelerating structure. The rf signals from the accelerating structure will be used in feedback control algorithm to calculate new rf pulse, so the signals measured in this case will be critical references for designing the algorithm. The field filling and dissipation processes can be monitored from the rf signals, which can help to characterize the accelerating structure. 

\begin{figure}[h]
\centering
\includegraphics[width=0.7\columnwidth]{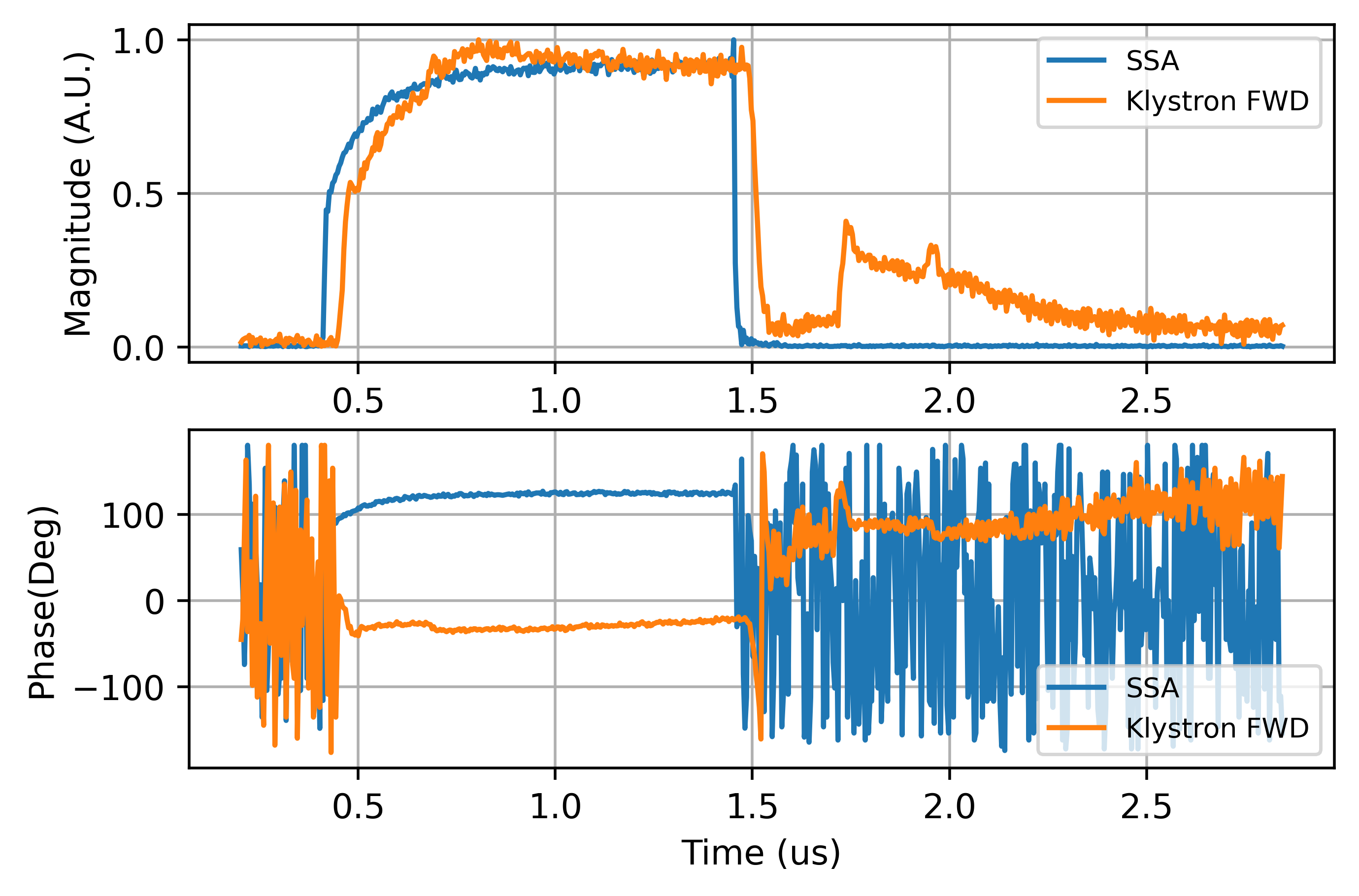}
\caption{SSA output and klystron forward measurements with peak power of 4.6 MW peak power at the klystron forward (DAC amplitude at 2000).}
\label{fig-5}       % Give a unique label
\end{figure}

\subsubsection{Klystron Forward Evaluation}

The high power analysis has been performed stage by stage and started with evaluating the output of klystron. These measurements were performed during the initial power conditioning of the accelerating structure. For rf pulse with duration of 1 \(\mu\)s, the structure was driven with a rf pulses at 10 Hz  within the desired breakdown rate of 100 per hour. Figure \ref{fig-5} shows the rf signals scaled to the maximum values at the first two stages of amplification, the SSA output signal and the klystron forward signal. The klystron introduces a delay of approximately 50 ns, but the waveform follows the SSA output. The signal rises again after the drive signal is down, which is due to the gird structure of rf power distribution. The klystron at Radiabeam supplies rf power for two separate lines, one connects to the C\(^3\) accelerating structure used for this test and the other connected to a separate test bunker for another test bunker. The rf power for klystron is divided by a power splitter and phase shifter (PSPS).  When the rf drive is turned off, the PSPS with the other rf line will reflect some power back to the klystron forward directional coupler and appeared as another peak on the pulse tail of the klystron forward signal captured in this case. The signal ramps down eventually as the rf power dissipate in the test setup.  The phase of the klystron forward signal changes when the input is ramping up, but settles when the input reached the the maximum and remains stable until the rf pulse is switched off.

\subsubsection{Accelerating Structure rf Measurements}

When the high power test is performed with the NG-LLRF platform, we are still conditioning the accelerating structure cavities. The peak power delivered at this point in the process is approximately 16.45 MW with pulse width around 450~ns at 60~Hz pulse rate with breakdown rate within the desired range. 

\begin{figure}[h]
\centering
\includegraphics[width=0.7\columnwidth]{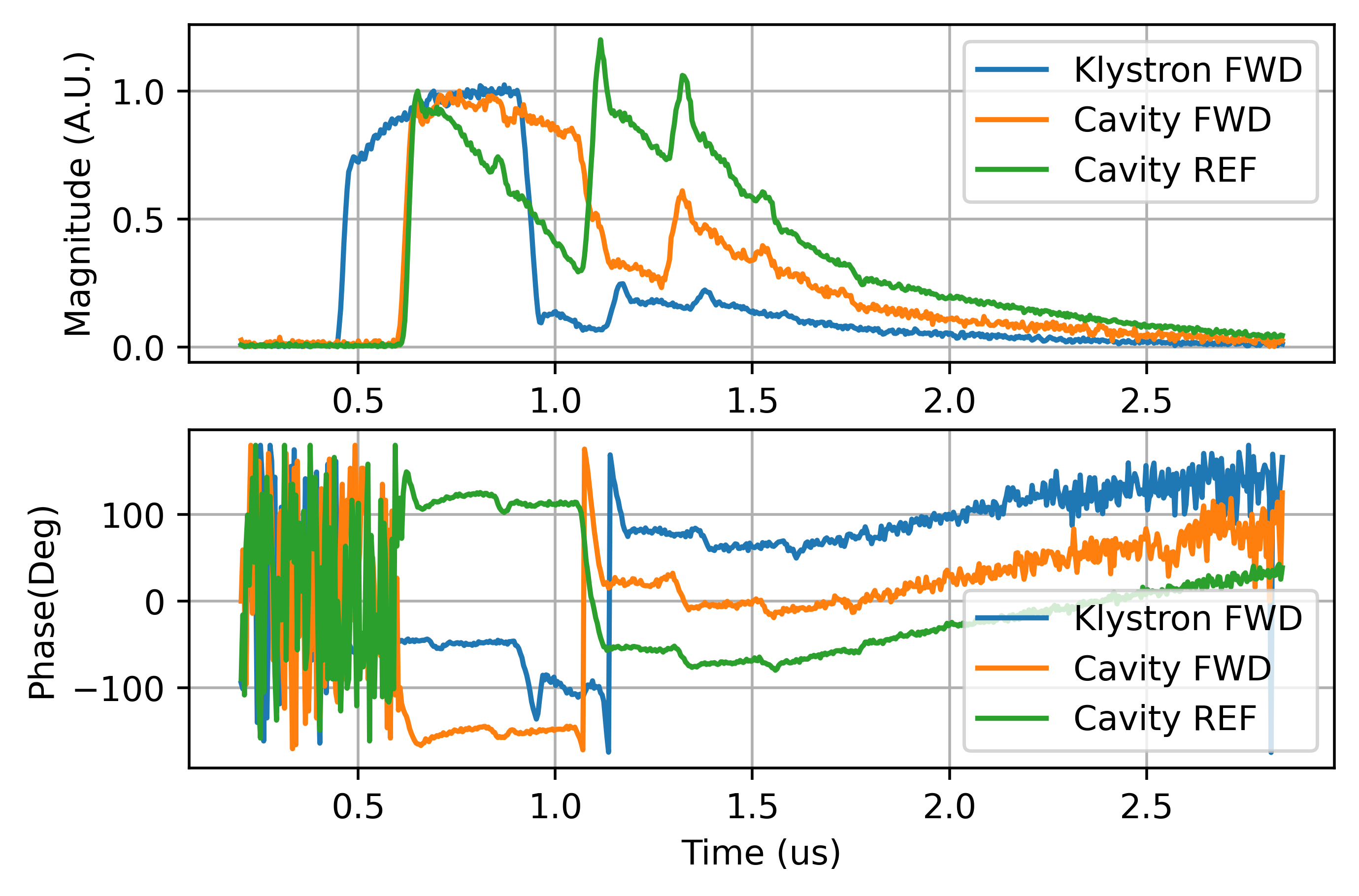}
\caption{ Klystron forward, cavity forward and cavity reflection measurements with a 450 ns rf pulse with peak power of 16.45 MW delivered to test bunker.}
\label{fig-6}       % Give a unique label
\end{figure}

In Figure \ref{fig-6}, the three rf signals, klystron forward, cavity forward and reflection, captured in parallel with the NG-LLRF are shown in the same plot. As the duration of rf pulse is only about 450 ns, the klsytron forward is down shortly after ramped up to the flat-top. The filling process is clearly reflected from the cavity signals. The cavity forward signal slowly reduces with time. The cavity reflection declines linearly with a steep slop as the rf field fills the cavity structure. Due to the short pulse duration, the magnitude of reflection signal does not reach zero, which means the filling process could not be completed before the rf shuts down. When the forward rf pulse is terminated, the stored energy radiates from the cavity resulting in the second peak of the reflected signal. The second peak of the reflected signal is large then the first which is characteristic of an overcoupled cavity. This cavity is overcoupled by design when no beam is present. With a beam, the cavity will be critically coupled. The phase of the reflection signal remains flat in the filling process, but is flipped by 180 degrees right after the rf power is turned down. The decay on the reflection signal after the second peak shows the rf power dissipation until it drops down to zero. Similar peaks appear after the rf power is turned off as seen on the klystron forward signal, the reflected rf power from PSPS and the other rf line is picked up at both the cavity forward and cavity refection couplers and appear as a peak while the signal is ramping down after the rf power is off.

\begin{figure}[h]
\centering
\includegraphics[width=0.7\columnwidth]{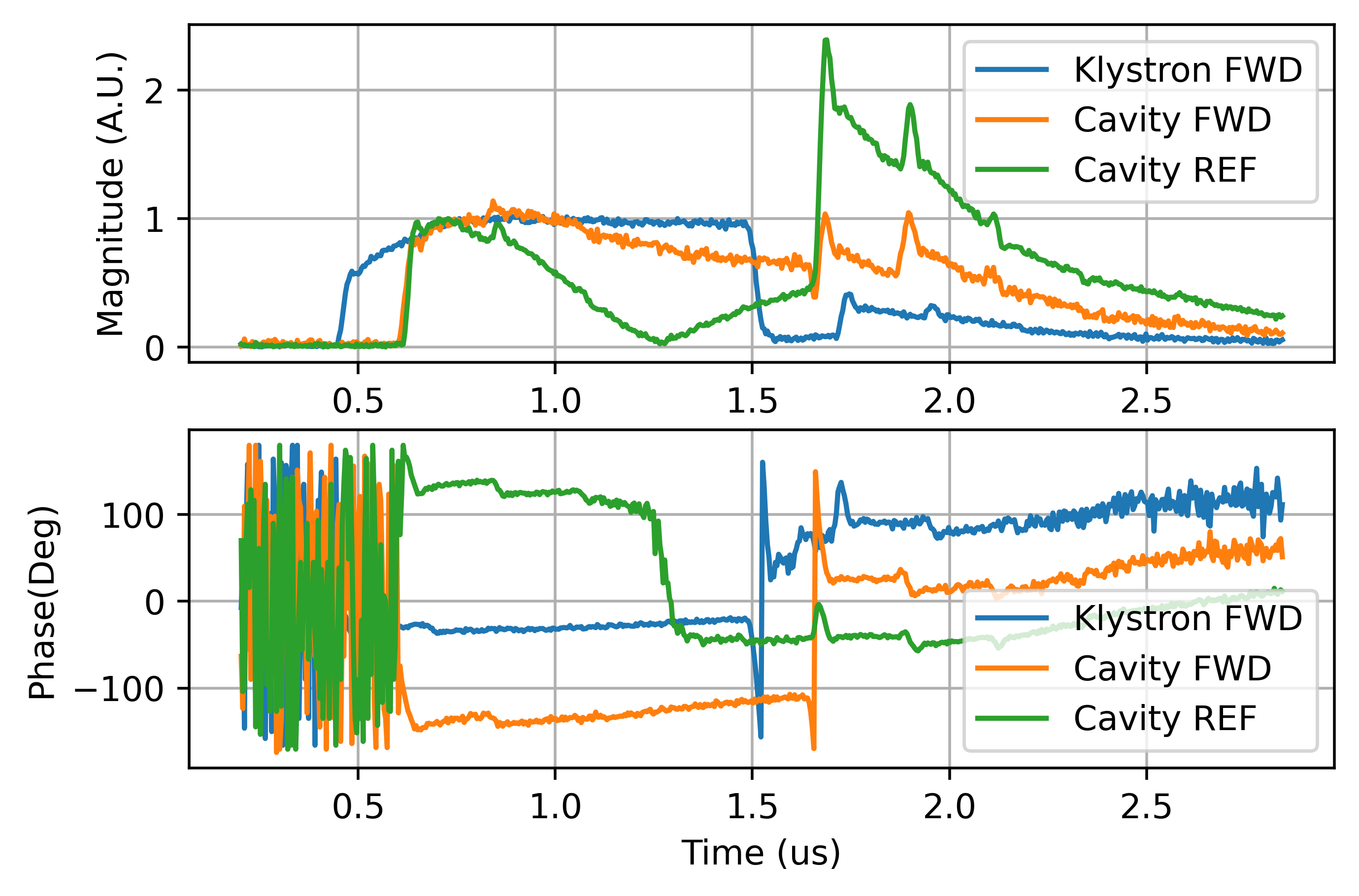}
\caption{ Klystron forward, cavity forward and cavity reflection measurements with a 1 \(\mu\)s rf pulse with peak power of 5.17 MW delivered to test bunker.}
\label{fig-7}       % Give a unique label
\end{figure}

For a 450 ns rf pulse, the accelerating structure could not be filled before the rf is turned off. The accelerating structure has also been tested with 1 \(\mu\)s rf pulse duration, but with 10 Hz pulse rate and 5.17 MW peak power. As Figure \ref{fig-7} shows, the klystron forward signal ramped up to a flat top with a duration over 500 ns. The magnitude of cavity reflection signal declines linearly with time and reaches zero about 650 ns after the rf power reached the structure, which remarks the accelerating structure filling process is completed. As rf power injected to the accelerator structure after it has been filled and the gradient of rf dissipation in the cavity wall is much lower than the injection, the excess rf power is reflected back. Therefore, the magnitude of cavity reflection start ramping up after the structure is filled, but the phase of cavity reflection signal flipped. 

\section{Conclusion}

The NG-LLRF platform developed at SLAC, including hardware, software and firmware, has been introduced in this paper. The prototype NG-LLRF system has been tested with the high power test stand with C\(^3\) accelerating structure. The rf performance at various stages of the test structure has been evaluated or analyzed. The SSA of the test setup delivered the rf pulse within the 150 fs phase jitter requirement of C\(^3\) at higher input power. The frequency domain analysis of SSA output shows a single dominant tone at the centre frequency, which is excellent as the drive for the klystron. From the rf signals captured from the klystron forward and accelerating structure couplers, the rf field filling process in the accelerating structure can be fully visualized. We have analyzed the rf signals at different pulse widths and power levels and all the features for the field filling process can be successfully resolved by the NG-LLRf platform. The rf signals from the accelerating structure are highly valuable for the LLRF feedback control firmware algorithms, which is under development. 

The experiments discussed in this paper focus on the NG-LLRF platform with the C\(^3\) structure. However, the RFSoC-based LLRF control platforms are also applicable to L-band and S-band accelerators with high stability levels. The bandwidth for the RFSoC devices is below the X-band range, but it can be used with analogue RF up and down mixing circuits and the integrated data converters and programmable logic in RFSoC can still offer advantages in the size, weight and power consumption.

\section{Acknowledgement}

The authors express their gratitude to Diego Amirari, Ronald Agustsson and Robert Berry from Radiabeam for their generous support of the high power test and  Ankur Dhar and Dennis Palmer for their insightful discussions, which have significantly contributed to this study. The work of the authors is supported by the U.S. Department of Energy under Contract No. DE-AC02-76SF00515.
% BibTeX or Biber users please use (the style is already called in the class, ensure that the "wocd.bst" style is in your local directory)
\bibliography{report} % Replace "your_bib_file" with the actual name of your .bib file

\end{document}